\documentclass[aps,prl,amsmath,amsfonts,amssymb,superscriptaddress,twocolumn,showpacs]{revtex4}
\renewcommand{\vec}[1]{\mathbf{#1}}
\usepackage{graphicx}
\usepackage{wasysym}
\usepackage{setspace}
\usepackage{color}
\usepackage{nicefrac}

\renewcommand{\Re}{\operatorname{Re}}
\renewcommand{\Im}{\operatorname{Im}}

\newcommand{\figref}[1]{Fig.~\ref{fig:#1}}
\newcommand{\Figref}[1]{Figure~\ref{fig:#1}}

\renewcommand{\eqref}[1]{Eq.~(\ref{eq:#1})}

\def\a{s}
\def\b{s}
\newcommand{\add}[1]{\if\a\b{{\color{red} #1}}\else{#1}\fi}
\newcommand{\comm}[1]{\if\a\b{{\color{blue}\{\small \sc #1\}}}\else{}\fi}
\newcommand{\del}[1]{{\if\a\b{{\color{magenta}[[#1]]}}\else{}\fi}}

\begin{document}

\title{Frequency-selective near-field radiative heat transfer between
  photonic-crystal slabs: A computational approach for arbitrary geometries and materials}

\author{Alejandro W. Rodriguez}
\affiliation{School of Engineering and Applied Sciences,
Harvard University, Cambridge, MA 02138}
\affiliation{Department of Mathematics, 
Massachusetts Institute of Technology, Cambridge, MA 02139}
\author{Ognjen Ilic}
\affiliation{Department of Physics,
Massachusetts Institute of Technology, Cambridge, MA 02139}
\author{Peter Bermel}
\affiliation{Department of Physics,
Massachusetts Institute of Technology, Cambridge, MA 02139}
\author{Ivan Celanovic}
\affiliation{Department of Physics,
Massachusetts Institute of Technology, Cambridge, MA 02139}
\author{John D. Joannopoulos}
\affiliation{Department of Physics,
Massachusetts Institute of Technology, Cambridge, MA 02139}
\author{Marin Solja{\v{c}}i{\'{c}}}
\affiliation{Department of Physics,
Massachusetts Institute of Technology, Cambridge, MA 02139}
\author{Steven G. Johnson}
\affiliation{Department of Mathematics,
Massachusetts Institute of Technology, Cambridge, MA 02139}

\begin{abstract}
  We demonstrate the possibility of achieving enhanced
  frequency-selective near-field radiative heat transfer between
  patterned (photonic crystal) slabs at designable frequencies and
  separations, exploiting a general numerical approach for computing
  heat transfer in arbitrary geometries and materials based on the
  finite-difference time-domain method.  Our simulations reveal a
  tradeoff between selectivity and near-field enhancement as the
  slab--slab separation decreases, with the patterned heat transfer
  eventually reducing to the unpatterned result multiplied by a fill
  factor (described by a standard proximity approximation).  We also
  find that heat transfer can be further enhanced at selective
  frequencies when the slabs are brought into a glide-symmetric
  configuration, a consequence of the degeneracies associated with the
  non-symmorphic symmetry group.
\end{abstract}

\maketitle

\begin{figure}[t]
\includegraphics[width=0.88\columnwidth]{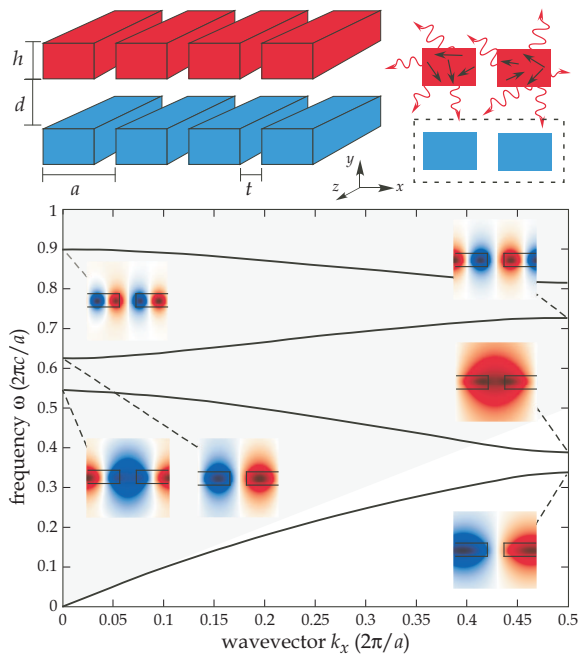}
\caption{Top: Schematic geometry of two PhC slabs of thickness
  $h=0.2a$, separated by a distance $d$, with periodic air grooves of
  period $a$ and width $t=0.2a$. Bottom: Modal frequencies
  $\omega(k_x)$ of the TE ($\vec{H} \cdot \vec{\hat{z}}=0$), $k_z=0$
  modes of an isolated slab; note the presence of both leaky (gray
  region) and guided modes. \emph{Insets}: mode profiles at $k_x =
  \{0, \pi/a\}$.}
\label{fig:fig1}
\end{figure}

Radiative transfer of energy from a hot to a cold body is well known
to be enhanced (even exceeding the black-body limit) when the bodies
are brought close enough for evanescent fields to contribute
flux~\cite{PolderVanHove71,Loomis1994,Pendry1999,Mulet02,Joulain05,Carey06,Volokitin07,Zhang07,BasuZhang09}. In
this paper, we demonstrate that near-field radiative transfer can be
greatly modified by using periodically patterned photonic-crystal
(PhC) surfaces, with frequency-selective enhancement that can be
controlled by choosing the geometry (rather than relying on material
or plasmon resonances available only at long
wavelengths~\cite{Francoeur08,FuTan09,Shen09}).  Until now,
investigations of near-field transfer in microstructured geometries
have been hampered by the lack of computational modelling techniques,
and we employ a new rigorous approach based on directly simulating
Maxwell's equations in time with random thermal sources, an extension
of the Langevin approach we previously used to model the emissivity of
a single body~\cite{Luo04:thermal}. Previously, aside from
semi-analytical results for planar
structures~\cite{Biehs2007,FuTan09,LauShen09}, formulations have been
developed that in principle handle arbitrary geometries but which thus
far have only been evaluated using Fourier methods specialized only
for pairs of spheres and/or
plates~\cite{narayanaswamy2008b,bimonte09,Wen10,Messina11,Kruger11,Otey11}. Multilayer
planar structures are tractable~\cite{BenAbdallah10}, but
(Fabry--Perot) resonant modes created in such structures are not
evanescent in the air gap, unlike leaky modes in transverse-patterned
structures (each of which has \emph{both} evanescent and propagating fields in the gap), so multilayer films do not lend themselves to
frequency-selective near-field enhancement. In the far field, it is
known that more complicated structures such as PhCs can be
designed to resonantly enhance radiative transfer at desired
frequencies~\cite{Greffet2002,Chan06,BermelGh10}, which is crucial for
applications such as thermophotovoltaic cells in which only certain
frequencies can be efficiently converted to
power~\cite{Whale02,Laroche06:tpv,BasuZhang09,BermelGh10}.  One would
like to obtain similar short-wavelength ($\lesssim
2\,\mu$m~\cite{BermelGh10}) enhancement of near-field effects, whereas
plasmon resonances are too long-wavelength for such applications.  In
a simple model system consisting of two PhC slabs, thin films with
periodic grooves [\figref{fig1}(top)], we show that the resonant leaky
modes created by the periodicity yield orders of magnitude enhancement
in the flux at designable wavelengths even for moderate separations
(100s of nm to microns for infrared wavelengths) compared to similar
structures in the far field, starting with weakly absorbing thin slabs
that transfer $< 1/1000$ of the flux between black bodies in the far
field. Furthermore, we show that the selective enhancement can be
almost doubled at selective peaks by using a glide-symmetric
configuration [\figref{fig3}(bottom)], due to degeneracies resulting
from the properties of the non-symmorphic symmetry
group~\cite{Mock10}. Ultimately, there is a tradeoff between
enhancement and frequency selectivity---much larger enhancement is
theoretically attained for~nm
separations~\cite{Mulet02,Joulain05,Volokitin07,Zhang07,BasuZhang09}
where geometric resonances have no effect, at the cost of frequency
selectivity and much more difficult fabrication---and this letter
offers a first glimpse of the practical design space that is available
to optimize these considerations.

Given two arbitrary bodies, at temperatures $T_1$ and $T_2$, their
radiative heat transfer $H(\omega,T_1,T_2)$ is:
\begin{equation}
  H(\omega; T_1, T_2) = \Phi(\omega) \left[\Theta(\omega, T_1) -
    \Theta(\omega,T_2)\right],
\label{eq:hPhi}
\end{equation}
where $\Phi$ is the flux into a single object due to random (white
noise) current sources present in the other object, and $\Theta$ is
the Planck distribution~\cite{BasuZhang09}. (Physically, for linear
electromagnetism it is equivalent to use Planck-distributed thermal
fluctuations or to use white-noise sources multiplied afterward by
$\Theta$.)  Computationally, this formulation can be expressed
directly in terms of a Langevin model~\cite{Luo04:thermal}, in which
white-noise sources are introduced into the evolution of Maxwell's
equations in a finite-difference time-domain (FDTD) method. (More
sophisticated approaches for solving this problem exist,
e.g. non-stochastic methods in the frequency domain, but the sacrifice
in efficiency of our FDTD approach is compensated by its generality
and simplicity.) Although we previously only applied this method to
equilibrium situations~\cite{Luo04:thermal}, the extension to
nonequilibrium situations is straightforward because the statistical
independence of random currents in different objects allows one to
calculate the flux due to thermal sources for one object at a time.
Reciprocity allows one to calculate the flux $\Phi$ from one body to
the other and infer the flux in the other direction, a fact that is
implicit in \eqref{hPhi}.

We investigate the proof-of-concept structure shown in
\figref{fig1}(top), involving two PhC dielectric slabs (thin films) of
thickness $h=0.2a$, separated by a distance $d$ in the $x$ direction,
with $z$-oriented air grooves of period $a$ and width $t=0.2a$ in the
$x$ direction. The permittivity of the slabs is taken to be of the
Drude form $\varepsilon(\omega) = \varepsilon_\infty - \sigma / \omega
(\omega + i \gamma)$, where $\varepsilon_\infty = 12.5$, $\sigma =
0.2533~(2\pi c/a)^2$, and $\gamma = 1.5915~(2\pi c/a)$, approximating
a dispersionless dielectric with $\Re \varepsilon \approx 12.5$ and
$\Im \varepsilon \approx 1$ at wavelengths comparable to~$a$. The
computation of $\Phi$ for this structure involves introducing
uncorrelated white-noise sources into the damped polarization
equation~\cite{Luo04:thermal} at each position (pixel) of one of the
bodies [red slab in \figref{fig1}(top)], and integrating the resulting
flux spectrum over a surface surrounding the other slab [blue slab in
  \figref{fig1}(top)].  The resulting spectrum is ensemble averaged
over many simulations ($\sim 60$) to reduce the noise level in the
spectrum. We employ a periodic unit cell in $x$ with Bloch-periodic
boundaries (phase difference $e^{ik_x a}$), truncate the cell in the
$y$ direction using standard PML boundary conditions, and integrate
$\Phi$ over $k_x$. The translation-invariant $z$ direction can
similarly be handled by integrating 2d simulations over $k_z$. We
first consider the purely 2d problem corresponding to $k_z = 0$ and
comment on the full 3d problem further below. At the moderately large
separations $\sim a$ studied here (assuming $a$ in the microns), the
TE modes ($\vec{H} \cdot \vec{\hat{z}}=0$) of the structure exhibit
stronger confinement than the TM modes~\cite{JoannopoulosJo08-book},
and end up dominating the heat transfer at all relevant~$\omega$, and
therefore we only compute the TE contribution.

\Figref{fig1} plots the (TE, $k_z=0$) modal frequencies $\omega(k_x)$
of a \emph{single} (isolated) slab. Modes that lie in the gray region
(above the $\omega=c k_x$ light line) are leaky
modes~\cite{JoannopoulosJo08-book}, which radiate into the air and
therefore contribute to heat transfer between the slabs in the far
field. Modes that lie below the light line are guided
modes~\cite{JoannopoulosJo08-book}, which evanescently decay in air
and thus transfer energy only in the near field.

\begin{figure}[t]
\includegraphics[width=0.87\columnwidth]{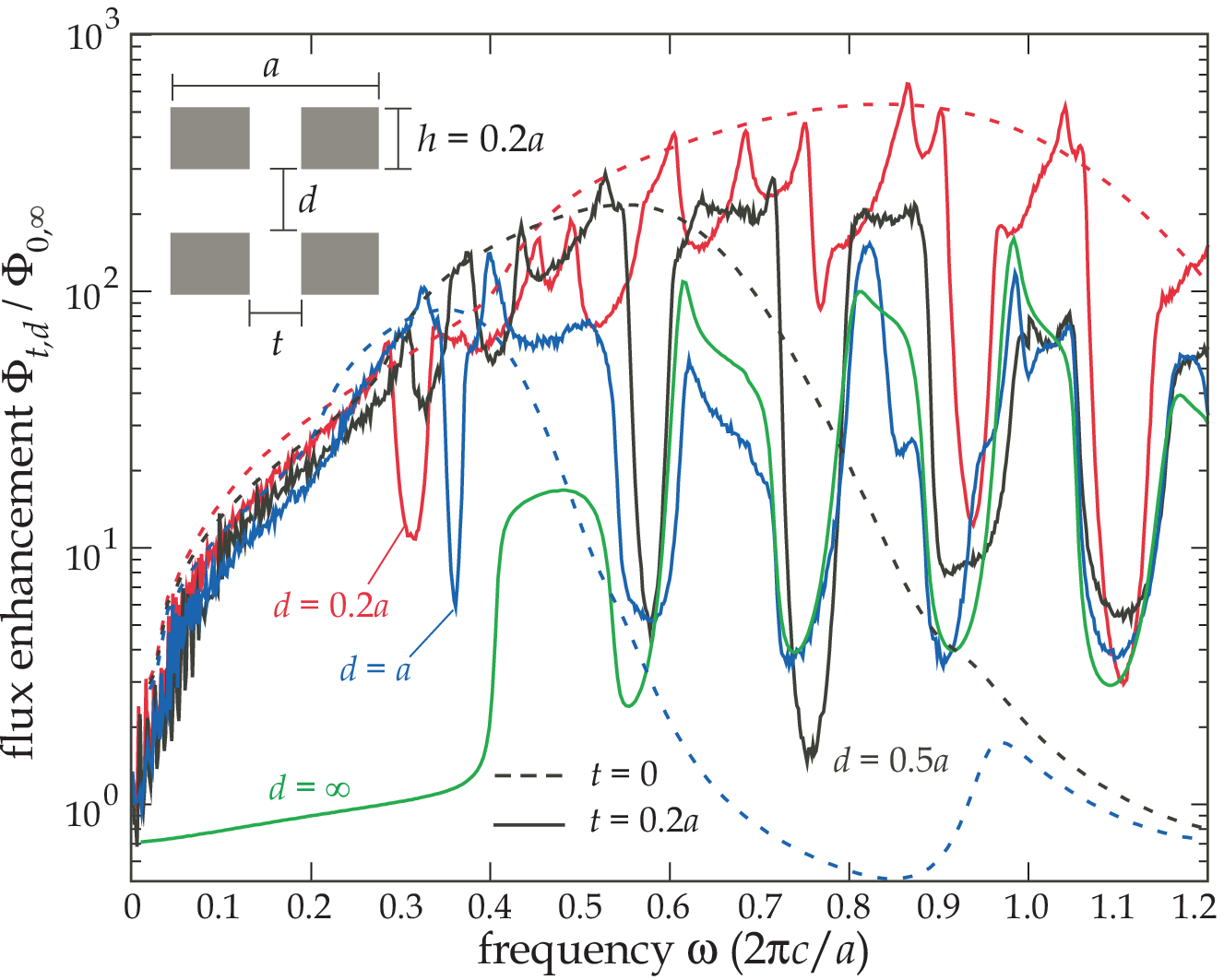}
\includegraphics[width=0.87\columnwidth]{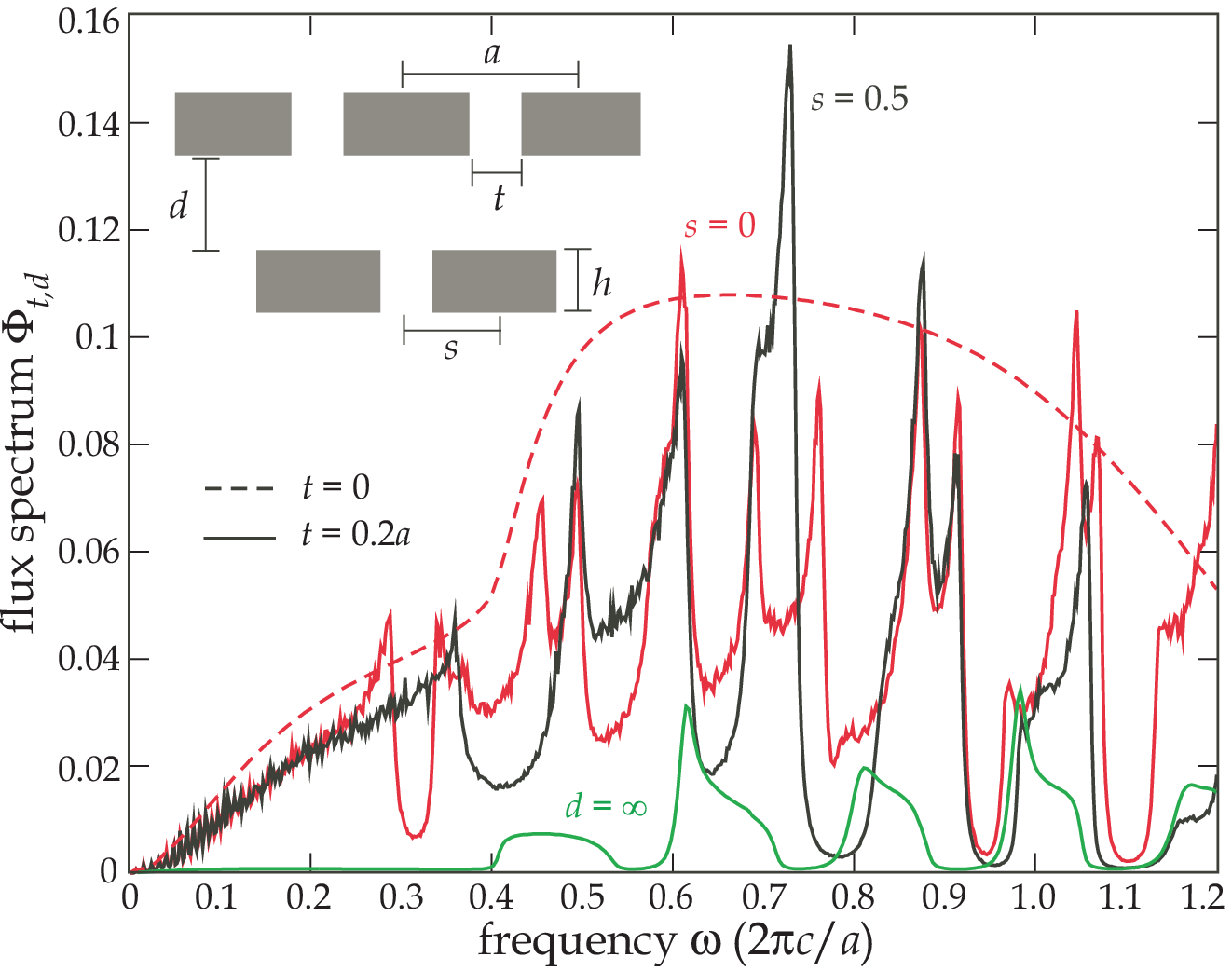}
\caption{Top: Flux enhancement $\eta = \Phi_{0.2a,d} /
  \Phi_{0,\infty}$ as a function of $\omega$ (units of $2\pi c/a$),
  where $\Phi_{t,d}$ is the flux spectrum between PhC slabs of groove
  width $t$ and separation $d$, at various $d = 0.2a, 0.5a, a,
  \infty$. Solid (dashed) lines denote $\eta$ for PhC (unpatterned)
  slabs.  Bottom: Flux spectrum $\Phi_{0.2a,0.2a}$ [units of $\hbar c
    (2\pi/a)^2$] for a symmetric ($s=0$, red line) or glide-symmetric
  ($s=0.5a$, black line) slab--slab configuration.}
\label{fig:fig2}
\end{figure}

\Figref{fig2}(top) plots the flux enhancement $\eta = \Phi_{t,d} /
\Phi_{0,\infty}$ as a function of $\omega$, where $\Phi_{t,d}$ is the
flux spectrum between PhC slabs of groove width $t$ and separated by
distance $d$, here normalized by the flux spectrum $\Phi_{0,\infty}$
of the unpatterned slabs ($t=0$) in the far field ($d \to \infty$) to
clearly illustrate the effect of finite $t$ and
$d$. ($\Phi_{0,\infty}$ is computed by Kirchoff's law from the
reflectivity in the far field~\cite{Mulet02}, by transfer-matrix
methods.) The $\Phi_{0,\infty}$ spectrum by itself is nearly flat, so
the spectral features in \figref{fig2} are not due to the
normalization.  Note that we are starting with very weakly absorbing
slabs (thin films), i.e. $\Phi_{0,\infty}$ is more than $1000$ times
smaller than the corresponding $\Phi=1$ of two black bodies.


Radiative heat transfer can be significantly changed by the
periodicity of the slabs~\cite{Chan06}, as illustrated by the green
curve in \figref{fig2}(top), which shows $\eta$ for PhCs of groove
width $t=0.2a$ in the far field. $\Phi_{0.2a,\infty}$ exhibits
wide-bandwidth peaks, orders of magnitude larger than
$\Phi_{0,\infty}$. Not surprisingly, the bandwidths of these peaks
match the bandwidths of the leaky-mode bands in \figref{fig1}(bottom),
whose integrated contribution over all $k_y$ leads to a smearing of
the sharp spectral peaks that occur at the leaky-mode frequencies for
each $k_y$.  The sharp drop in $\eta$ between the peaks is a
consequence of the presence of pseudogaps between the leaky-mode
bands.


As the separation decreases, a number of dramatic features are
observed in \figref{fig2}(top), which shows $\eta$ at three additional
separations ($d=0.2a$, $d=0.5a$ and $d=a$).  At $d=a$, one can observe
a dramatic increase in $\eta$ at low frequencies. This arises due to
the evanescent coupling and contribution of guided modes to $h$.  This
near-field coupling increases further at smaller $d$, and additionally
(especially in the $d=0.2a$ curve) one can see a dip corresponding to
the gap in the guided modes of \figref{fig1}(bottom), and sharp peaks
at the edges of the gap (a consequence of van Hove
singularities~\cite{JoannopoulosJo08-book}).  There is also an
enhancement of $\Phi$ at small $d$ at the leaky-mode peaks because, in
a patterned slab, leaky-modes have an evanescent as well as a
radiative component~\cite{JoannopoulosJo08-book}, and the former
enhances the coupling at short distances. As for the guided modes,
this evanescent coupling is further enhanced by the van Hove
singularities of the zero-slope ``slow-light'' band edges.
Also, the leaky modes split into bonding and antibonding pairs
as $d$ decreases, and a similar splitting is visible in the guided
mode peaks at even smaller $d$, a phenomenon visible even more clearly
in \figref{fig2}(bottom).

A striking phenomenon in the near-field interaction occurs when one
patterned slab is shifted in the $x$ direction by $0.5a$ with respect
to the other slab, as shown in \figref{fig2}(bottom).  In this case
(plotted here as absolute $\Phi$ on a linear scale), the
bonding-antibonding peaks of the symmetric configuration (solid red
curve) \emph{merge} at certain $\omega$, and can even add to
approximately double the heat transfer of the unpatterned slab.  This
arises due to the glide symmetry of the shifted structure, whose
non-symmorphic symmetry group supports degeneracies at the Brillouin
zone edge ($k_x=\pi/a$)~\cite{Mock10}, causing the corresponding
bonding-antibonding mode splittings to disappear.  In contrast, the
frequency splitting corresponding to the $k_x=0$ band edges, e.g. at
$\omega \approx 0.9~(2\pi c/a)$, persists even for the glide-symmetric
structure.  Similarly, the gap in the guided modes of
\figref{fig1}(bottom), near $\omega \approx 0.4~(2 \pi c/a)$, is
greatly modified by the presence of the other slab in the
mirror-symmetric case (red curve) but not in the glide-symmetric case
(black curve).


\Figref{fig2}(top) also shows the enhancement for unpatterned ($t=0$)
slabs.  These also exhibit a striking near-field enhancement for $d <
a$, but in this case the enhancement is broad-band, because there are
no standing-wave leaky-mode solutions in an unpatterned slab to induce
frequency selectivity.  In particular, at relatively large separations
(e.g. $d=a$), the patterned slab exhibits more than 10 times the
enhancement of the unpatterned slab, due to these resonant
contributions.  At $d=0.5a$, the unpatterned slab is better for small
$\omega$ and is worse for larger $\omega$, where the latter is due to
the fact that the evanescent interactions become shorter-range at high
$\omega$ and vanish in the absence of resonant enhancement.  At
$d=0.2a$, the unpatterned slab has larger enhancement over the whole
bandwidth shown here, but lacks frequency selectivity. In general, as
$d$ decreases, the periodic structure of the PhC slabs ceases to
matter, and the enhancement of the peaks relative to the overall
near-field enhancement is lost. At even smaller $d$, not shown here,
the interaction between the PhCs can be described by a proximity
approximation in which adjacent surfaces are treated as parallel
plates~\cite{Joulain05,Volokitin07,Zhang07,BasuZhang09} multiplied by
a fill factor.

\Figref{fig3} plots the total (integrated) heat transfer $H_{t,d}$ as
a function of separation $d$ for both patterned ($H_{t,d}$, squares)
and unpatterned ($H_{0,d}$, thin lines) slabs of period $a=2\mu$m, at
various temperatures $T_2$, and $T_1=0$. At $T_2 = 500$~K, dominated
by low-$\omega$ contributions, the unpatterned slabs yield larger $H$
over almost every $d$ (even as $d \to \infty$), which increases for
small~$d$ due to near-field enhancement. Moreover, we find that $\xi =
H_{t,d}/H_{0,d} \to 1-t/a = 0.8$ as $d \to 0$, as expected from the
proximity approximation.  (For TE modes, $H$ asymptotes to a finite
value as $d\to 0$, whereas for TM modes, whose contributions are
negligible at the separations we consider here, $H$ eventually
diverges~\cite{BasuZhang09}.)  At $T_2 = 1000$~K, larger $\omega$
begin to dominate, and the presence of PhC peaks [\figref{fig2}]
causes a significant increase in $H$. In the (purely radiative) limit
$d \to \infty$, the PhC slabs perform much better than the unpatterned
slabs ($\xi \sim 3$), although the tradeoff between enhancement and
selectivity is apparent as there exist a $d_c \approx 7~\mu$m below
which $\xi < 1$ (eventually $\to 0.8$).  As expected, $d_c$ decreases
as $T_2$ increases due to the increasing role of higher-frequency
resonances. At $T_2 = 2000$~K, we find that $\xi \approx 5$ at larger
$d$, increases at intermediate $d$ and falls below~1 at $d_c \approx
2~\mu$m.

\Figref{fig3} illustrates the advantages of using PhCs at large and
intermediate $d$, where both selectivity and near-field effects
coexist. However, in many applications, the figure of merit is not the
overall $H$ but the integrated $H$ over a \emph{finite frequency
  bandwidth}. In thermophotovoltaics, heat transferred from a hot body
to a cold semiconductor is converted to electricity via electronic
bandgap transitions, which means that photons of frequency smaller
than the bandgap frequency $\omega_g$ (bandgap wavelength of a few
microns) are lost in the conversion process~\cite{BasuZhang09}. The
left inset of \figref{fig3} shows a corresponding \emph{ad hoc} figure
of merit: the \emph{windowed} heat transfer $H_w = \int d\omega \,
H(\omega, T_1,T_2) n(\omega)$ between slabs of period $a=1\mu$m, as a
function of separation $d$, and at temperatures $T_1 = 300$~K and $T_2 =
1500$~K. The window function $n$ is similar to the ``quantum
efficiency'' function of a typical semiconductor of bandgap wavelength
$2\pi c/\omega_g = 2.5~\mu$m~\cite{BasuZhang09}. The spectra
$H(\omega)$ and $n(\omega)$ are plotted on the right inset of
\figref{fig3} at $d=2~\mu$m.  Noticeably, the presence of
high-frequency peaks in the PhC makes its $H_w$ over an order of
magnitude larger than that of the unpatterned slab for $d \gtrsim
1\,\mu$m; the situation is even more favorable for the PhC because the
low-frequency photons in the unpatterned case are wasted.

\begin{figure}[t]
\includegraphics[width=1.0\columnwidth]{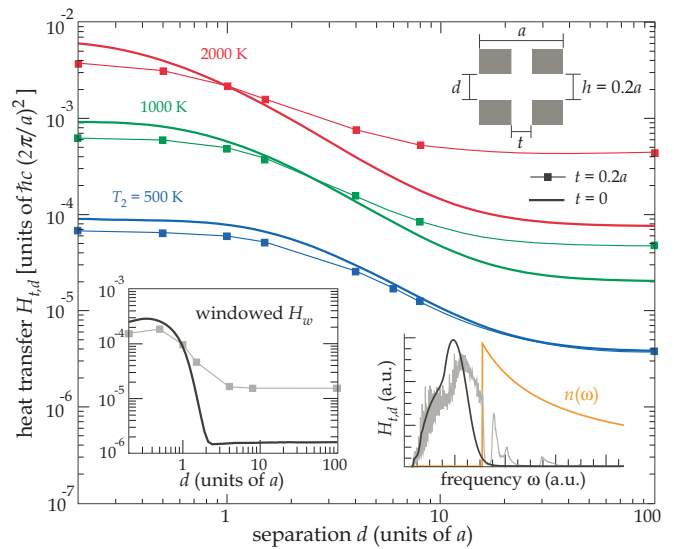}
\caption{Heat transfer $H_{t,d}$ [units of $\hbar c (2\pi /a)^2$]
  between unpatterned ($t=0$, lines) or PhC ($t=0.2a$, squares) slabs
  of period $a=2\mu$m, as a function of $d$ (units of $a$), plotted
  for $T_1=300$~K and various $T_2$. Left inset: $h_w$ as a function
  of $d$ (units of $a$) for slabs of period $a=1\mu$m, and $T_2 =
  1500$~K, where $H_w$ is computed by windowing the spectrum
  $h(\omega)$ with a function that excludes contributions of
  wavelengths $> 2.5\,\mu$m. Right inset: $H_{t,d}$ (gray) and
  $H_{0,d}$ (black) at a typical separation $d=2~\mu$m, and the window
  function $n(\omega)$ (orange line).}
\label{fig:fig3}
\end{figure}


Our calculations of $H$ were restricted to $k_z=0$ for computational
convenience. However, preliminary calculations of the $k_z$-integrated
$H(\omega)$, for a single separation, reveal no qualitative changes in
the spectrum, except for a slight broadening of the spectral peaks. As
noted above, the geometry of \figref{fig1} is in no way optimal and
represents only a proof of concept: we believe that a 2d-periodic
geometry will yield even better performance, an exciting direction for
future numerical studies.


This work was supported by DARPA contract N66001-09-1-2070-DOD (AWR),
by S3TEC, an Energy Frontier Research Center funded by the US DOE,
Office of Science, and Office of Basic Energy Sciences, under award
No. DE-SC0001299, by the US ARO contract W911NF-07-D-0004, and through
the Army Research Office through the Institute for Soldier
Nanotechnologies under Contract No. W911NF-07-D0004 (OI, IC, PB).


\end{document}